\documentclass[twocolumn,prl,superscriptaddress]{revtex4-2}
\usepackage{amsfonts}
\usepackage{amsmath}
\usepackage{amssymb}
\usepackage{mathrsfs}
\usepackage{graphicx}
\usepackage{txfonts}
\usepackage{xcolor}%
\usepackage{ulem}
\usepackage{appendix}
\usepackage{changes}
\usepackage{url}
\usepackage[colorlinks=true, citecolor=blue, urlcolor=blue]{hyperref}
\providecommand{\U}[1]{\protect\rule{.1in}{.1in}}

\begin{document}
\title{Loophole-free test of macroscopic realism via high-order correlations of measurement}
\today

\author{Ping Wang}
\email{wpking@bnu.edu.cn}
\affiliation{College of Education for the Future, Beijing Normal University, Zhuhai 519087, China}
\affiliation{Department of Physics, The Chinese University of Hong Kong, Shatin, New Territories, Hong Kong, China}
\affiliation{Centre for Quantum Coherence,
The Chinese University of Hong Kong, Shatin, New Territories, Hong Kong, China}
\affiliation{The Hong Kong Institute of Quantum Information Science and Technology,
The Chinese University of Hong Kong, Shatin, New Territories, Hong Kong, China}

\author{Chong Chen}
\affiliation{Department of Physics, The Chinese University of Hong Kong, Shatin, New Territories, Hong Kong, China}
\affiliation{Centre for Quantum Coherence,
The Chinese University of Hong Kong, Shatin, New Territories, Hong Kong, China}
\affiliation{The Hong Kong Institute of Quantum Information Science and Technology,
The Chinese University of Hong Kong, Shatin, New Territories, Hong Kong, China}

\author{Hao Liao}
\affiliation{National Engineering Laboratory on Big Data System Computing Technology, Guangdong Province Engineering Center of
China-made High Performance Data Computing System, College of Computer Science and Software Engineering, Shenzhen
University, Shenzhen 518060, China}

\author{Vadim V. Vorobyov}
\affiliation{3rd Institute of Physics, Research Center SCoPE and IQST, University of Stuttgart, 70569 Stuttgart, Germany}

\author{ J\"org Wrachtrup}
\affiliation{3rd Institute of Physics, Research Center SCoPE and IQST, University of Stuttgart, 70569 Stuttgart, Germany}

\author{Ren-Bao Liu}
\email{rbliu@cuhk.edu.hk}
\affiliation{Department of Physics, The Chinese University of Hong Kong, Shatin, New Territories, Hong Kong, China}
\affiliation{Centre for Quantum Coherence,
The Chinese University of Hong Kong, Shatin, New Territories, Hong Kong, China}
\affiliation{The Hong Kong Institute of Quantum Information Science and Technology,
The Chinese University of Hong Kong, Shatin, New Territories, Hong Kong, China}
\affiliation{New Cornerstone Science Laboratory,
The Chinese University of Hong Kong, Shatin, New Territories, Hong Kong, China}

\begin{abstract}
Test of {macroscopic realism} (MR) is key to understanding
the foundation of quantum mechanics. Due to the existence of the  {non-invasive measurability} loophole and other interpretation loopholes, however, such test remains an open question. Here we propose a general inequality based on high-order correlations of measurements for a loophole-free test of MR at the weak signal limit. Importantly, the inequality is established using the statistics of \textit{raw data} recorded by classical devices, without requiring a specific model for the measurement process, so its violation would falsify MR without the interpretation loophole. The non-invasive measurability loophole is also closed, since the weak signal limit can be verified solely by measurement data (using the relative scaling behaviors of different orders of correlations). We demonstrate that the inequality can be broken by a quantum spin model. The inequality proposed here provides an unambiguous test of the MR principle and is also useful to characterizing {quantum coherence}. 
\end{abstract}

\maketitle


\textit{Introduction.}  The foundation of quantum
mechanics has been under examination since its birth. The Bell inequality~\cite{BellPPF1964}
based on {local realism} was proposed to test the completeness
of quantum mechanics~\cite{BrunnerRMP2014,HensenNature2015,GiustinaPRL2015}.
While the test by the Bell inequality requires measurements with space-like
separation to exclude the communication loophole~\cite{ClauserRPP1978},
Leggett and Garg proposed an alternative protocol to
test the {macroscopic realism} (MR) without the need of considering special relativity~\cite{LeggettPRL1985}.
The MR states that a measurement on a macroscopic
system reveals a well-defined, {pre-existing} value~\cite{LeggettPRL1985}, that is,
a value that does not depend on how the quantity is measured.
The MR leads to a set of inequalities between the correlations of measurements
at different times, called Leggett-Garg inequalities, which can be broken in quantum
mechanics where the MR does not hold. The Leggett-Garg inequalities~\cite{LeggettPRL1985} have received
intensive attention~\cite{EmaryRPP2013} since their violation excludes
any classical theories bearing the feature of {realism} with, 
the {local realism} being a special case~\cite{PottsPRL2019}.
Moreover, the Leggett-Garg inequalities are also useful in characterizing macroscopic
quantum coherence and identifying quantum resources for quantum information technologies~\cite{FroewisRMP2018}. 

The test of Leggett-Garg inequalities need to avoid the
so-called clumsiness loopholes~\cite{WildeFOP2012,EmaryRPP2013}, especially the one related to the assumption of {non-invasive measurability} (NIM)~\cite{LeggettPRL1985},
which states that it is possible to measure an observable without
perturbing it if the MR principle holds. However, it is difficult to verify the NIM. Therefore, the violation of Leggett-Garg inequalities may result from unwitting invasivity of
the measurements rather than the absence of MR. A series of investigations attempted to exclude this loophole through quantification
of the invasivity of the measurements by controllable experiments~\cite{HuelgaPRA1995,KneeNC2012,RobensPRX2015,KneeNC2016}, which, 
however, introduced other assumptions~\cite{KneeNC2012,KneeNC2016}.
Another type of loopholes are related to the requirement of underlying measurement models to interpret the raw data. Since the outputs of classical measurement apparatuses always have positive probability distributions~\cite{DiracPRSL1942,CalarcoEPL1999},
the statistics of experimental raw data would never violate the Leggett-Garg inequalities.
Instead, one needs to extract the underlying values of physical variables from the raw data and
construct correlations of these underlying variables to test the MR; this procedure inevitably
relies on the knowledge of the specific model of the measurement process~\cite{RuskovPRL2006,PalaciosNP2010}, hence the interpretation loophole. Due to these loopholes, the test of MR remains
an open question despite many endeavors in various physical systems~\cite{PalaciosNP2010,GogginPNAS2011,WaldherrPRL2011,SouzaNJP2011,AthalyePRL2011,SuzukiNJP2012,KneeNC2012,GroenPRL2013,GeorgePNAS2013,RobensPRX2015,BechtoldPRL2016,FormaggioPRL2016,BednorzPRL2010,BednorzNJP2012}.

In this paper, we establish general and unambiguous inequalities between correlations in \textit{raw data} obtained at the weak signal limit (WSL)~\cite{WangPRL2019}, for a loophole-free
test of MR. The WSL condition can be independently verified only using the raw data. These inequalities
do not require additional assumptions such as NIM~\cite{PalaciosNP2010,GogginPNAS2011,WaldherrPRL2011,SouzaNJP2011,AthalyePRL2011,SuzukiNJP2012,KneeNC2012,GroenPRL2013,GeorgePNAS2013,RobensPRX2015,BechtoldPRL2016,FormaggioPRL2016},
specific measurement models~\cite{PalaciosNP2010,KneeNC2012}, or knowledge of dynamics~\cite{WaldherrPRL2011}.
Assuming only MR and the (independently verifiable) WSL, we construct
a general constraint on the relation between fourth- and second-order correlations of raw data. The violation of this constraint
would unambiguously falsify MR without an aforementioned loophole.
An example of violation is illustrated in a quantum spin model~\cite{BlokNatPhys2014,PfenderNC2019,PanNP2020,JonasNC2022}.

\textit{General theory.}  In a world complied
with MR, the state of a system at any time is
fully described by the probability of a set of variables taking certain \textit{pre-existing} values. In classical mechanics,
such variables can be pairs of canonical positions and momenta and
the probability is the Liouville density in phase space. The existence of
preexisting probability is also assumed for hidden-variable theories~\cite{BohmPR1952} trying to maintain the local realism in quantum theory.

The key to testing MR is the violation of
the positivity of the probability. For this goal, we resort to higher-order correlations of
data obtained with variable measurement strength. With only the assumption that the probability for the measurement apparatuses and the target system taking a specific state is positive (that is, the assumption of MR), we establish an inequality between the fourth- and second-order correlations 
of raw data obtained at the WSL (which itself can be checked using the relative strengths of the correlations in different orders).

\begin{figure}
\includegraphics[width=0.9\columnwidth]{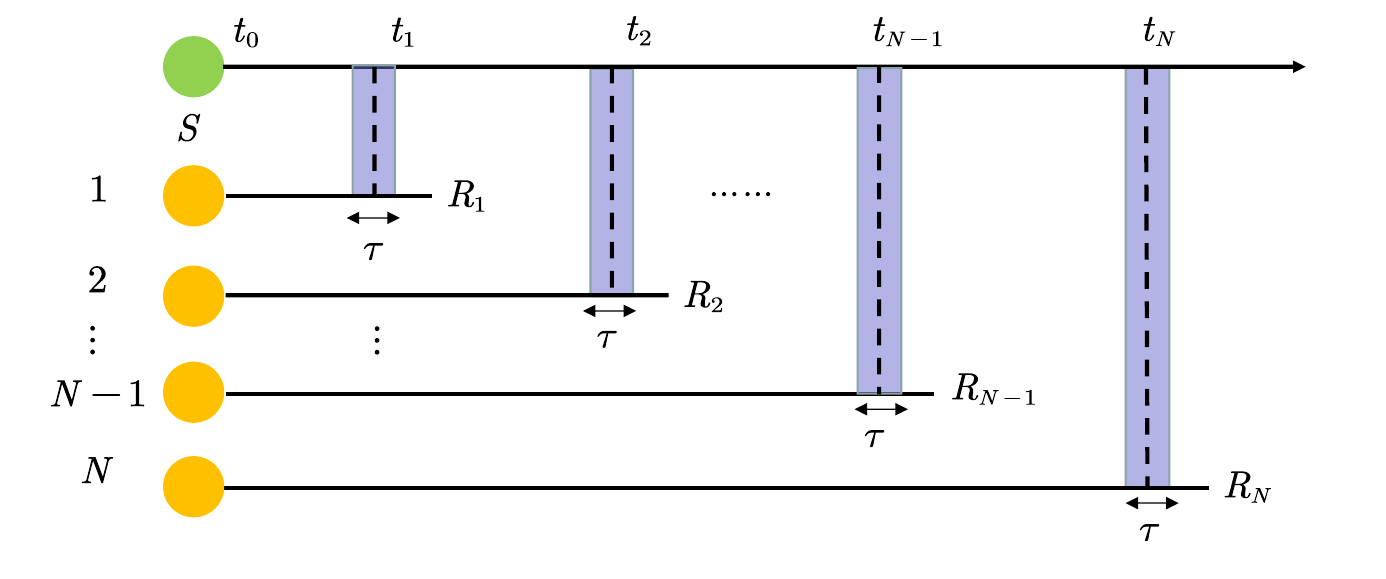}
\caption{\textbf{Schemes for measuring the correlations.} Consecutive
measurement in a unit sequence. The green circle denotes the target and the $N$ orange
circles denote $N$ apparatuses, which are in contact with
the target consecutively at times $t_{1}<t_{2}<\cdot\cdot\cdot<t_{N}$, each with
a duration $\tau$ (denoted by the dashed lines and the purple box). The outputs $R_{1},R_{2},\ldots,R_{N}$
are recorded to calculate the correlations.\label{fig:Fig1}}
\end{figure}

To construct the inequality, we examine the properties of the measurement
process when MR holds. We consider the measurement process shown
in Fig.~\ref{fig:Fig1}. At the initial time $t_{0}$, the target
and $N$ apparatuses are independent of each other. We introduce $s(t)$ and $r_j(t)$
to denote the values of the target and the $j$-th apparatus at time $t$, respectively. 
The $j$-th apparatus contacts the target at the time $t_j-\tau$
for a duration $\tau$. The WSL corresponds to $\tau\rightarrow 0$. The output of the $j$-th apparatus is recorded at time $t_j$ 
as $R_j=r_j(t_j)$. The outputs of $N$ consecutive measurements are denoted  $\mathbf{R}^{(N)}=\{R_1,R_2,\ldots,R_N\}$. After $N$ measurements, a  joint probability $P\left(\mathbf{R}^{(N)},S\right)$ for the initial value
$S\equiv s(t_0)$ of the target and the $N$ outputs is formed. 
We can always factorize the joined probability $P\left(\mathbf{R}^{(N)},S\right)$ into the form
\begin{equation}
P\left(\mathbf{R}^{(N)},S\right)=\prod_{n=1}^{N}P_n\left(R_n|\mathbf{R}^{(n-1)},S\right)P_{0}(S)\label{eq:joined dis}
\end{equation}
by autoregressive approach, where  $P_n\left(R_n\vert \mathbf{R}^{(n-1)},S\right)$
is the probability distribution of the $n$-th measurement output conditioned on the earlier measurement outputs and the initial state of the target system, and $P_0(S)$ is the probability distribution of the initial value of the target (before any measurement).

The key property of the conditional probability
$P_n\left(R_n\vert \mathbf{R}^{(n-1)},S\right)$ at the WSL is
\begin{equation}
P_n\left(R_n|\mathbf{R}^{(n-1)},S\right)=P_n(R_n)+\tau  D_n(R_n|S)+O(\tau ^2), \label{eq:conditionprob2}
\end{equation}
where $P_n(R_n)$ is the distribution of the output of the $n$-th apparatus without interacting with the target. 
Physically, $\tau D_n(R_n\vert S)$ quantifies how much information
can be extracted from the target in the $n$-th measurement in the leading order of $\tau$. Note that $D_n\left(R_N|S\right)$ is not positive-definite since $\int D_n\left(R_n|S\right)dR_n=0$ up to $O\left(\tau \right)$.
Equation~(\ref{eq:conditionprob2}) means, up to the first order of $\tau $, the $n$-th output is independent of the previous measurements, which is intuitive since the perturbation of the system due to the weak measurements at earlier times should vanish with $\tau \rightarrow 0$. The proof of the equation can be found in~\cite{supp}.

With Eq.~(\ref{eq:conditionprob2}), the joint probability of the outputs 
and the initial state of the target
can be written as
\begin{equation}
P\left(\mathbf{R}^{(N)},S\right)=\prod_{n=1}^{N}\left[P_{n}\left(R_n\right)+\tau  D_n\left(R_n|S\right)+O\left(\tau ^{2}\right)\right]P_{0}(S) .\label{eq:joint_prob}
\end{equation}
The WSL condition in Eq.~(\ref{eq:joint_prob}) 
is in sharp contrast to the NIM principle in Ref.~\cite{LeggettPRL1985}.
The NIM principle assumes that it is possible to measure a classical
quantity without perturbing it, which is difficult to check without
further assumptions; the WSL condition does not focus on how much
the target is perturbed but on how weak the signal is extracted from
the target, which can be tested experimentally.

In the following, we give a general form of the correlations of the measurement outputs in the WSL using Eq.~(\ref{eq:joint_prob}). 
We define the $L$-th order correlations of the $N$ measurements as
\begin{align}
C_{n_1,n_2,\ldots,n_L} = & \int \delta {R}_{n_1}\delta{R}_{n_2} \cdots  \delta {R}_{n_L} P\left({\mathbf R}^{(N)},S\right) d^N {\mathbf R}^{(N)}dS,
\label{eq:correlation_of_raw_data}
\end{align}
where ${n_1,n_2,\ldots,n_L}\in \{1,\cdot\cdot\cdot,N\}$ and $\delta R_j\equiv R_j-\left\langle R_j\right\rangle$ with $\left\langle R_j\right\rangle\equiv \int  R_j P\left({\mathbf R}^{(N)},S\right)d^N{\mathbf R}^{(N)}dS$
being the average of the $j$-th measurement. This definition removes the background correlations between the apparatuses and singles out the correlations established by the interaction between the apparatuses and the target system. Using 
Eq.~(\ref{eq:joint_prob}), we obtain the correlations as
\begin{align}
C_{n_1,n_2,\ldots,n_L} = \tau^L \left\langle\delta \bar{R}'_{n_1}\delta \bar{R}'_{n_2}\cdots\delta \bar{R}'_{n_L} \right\rangle_S + O\left(\tau^{L+1}\right),
\label{eq:correlations}
\end{align}
where $\left\langle F \right\rangle_S\equiv \int F P_0(S) dS$ is the average of $F$ over the distribution of the target state and $\delta \bar{R}'_j\equiv \bar{R}'_j-\left\langle\bar{R}'_j\right\rangle_S$ with $\bar{R}'_j\equiv \int R_j  D_j\left(R_j|S\right) dR_j$, the $j$-th output averaged over the probability distribution variation induced by the interaction with the target in the state $S$. 
Importantly, the correlations in Eq.~(\ref{eq:correlations}) have the form as an average over an \textit{a prior} positive probability distribution $P_0(S)$ (hence determined by MR). Therefore, we can establish constraints on relations between correlations of different orders using the following inequality 
\begin{equation}
\left|\left\langle F\right\rangle_S\right|^{2}\le\left\langle\left|F^{2}\right|\right\rangle_S,\label{eq:identity}
\end{equation}
which hold for any quantity $F$ that has a positive probability distribution $P_0(S)$. 

We consider the correlations for four measurements at $t_{i}<t_{j}<t_{k}<t_{l}$ and define the quantity of Clauser–Horne–Shimony–Holt type~\cite{BrunnerRMP2014}
\[
F=\left(\delta\bar{R}'_{i}+\delta\bar{R}'_{k}\right)\delta\bar{R}'_{j}+\left(\delta\bar{R}'_{i}-\delta\bar{R}'_{k}\right)\delta\bar{R}'_{l}.
\]
From Eq.~(\ref{eq:identity}), we obtain an inequality between the second-order correlations and the fourth ones
\begin{eqnarray}
 \tau ^{4}\left|\left\langle\left[\left(\delta\bar{R}'_i+\delta\bar{R}'_k\right)\delta\bar{R}'_j+\left(\delta\bar{R}'_i-\delta\bar{R}'_k\right)\delta\bar{R}'_l\right]\right\rangle_S\right|^2 & \le\nonumber \\
 \tau ^{4}{\left\langle\left|\left[\left(\delta\bar{R}'_i+\delta\bar{R}'_k\right)\delta\bar{R}'_j+\left(\delta\bar{R}'_i-\delta\bar{R}'_k\right)\delta\bar{R}'_l\right]^{2}\right|\right\rangle_S} & +O\left(\tau^5\right).
\label{Eq:inequalities}
\end{eqnarray}
{Note that the inequality still holds if there are extra measurements performed between the times $t_i, t_j, t_k$ and $t_l$ as long as the WSL condition ($\tau \rightarrow 0$) is satisfied.}

To establish an equality between correlations of raw data, we should replace two sides of Eq.~(\ref{Eq:inequalities}) with observable
signals. The right-hand side of Eq.~(\ref{Eq:inequalities}) contains equal-time
correlations such as $\left\langle\left( \delta\bar{R}'_j\right)^{2}\left(\delta\bar{R}'_k\right)^{2}\right\rangle$, which are not directly observed quantities. {To construct an inequality with
only observable signals, we use the fact that in the WSL $D_{j_+}(R|S)=D_j(R|S)+O\left(\tau\right)$ and therefore $\left(\delta\bar{R}'_{j}\right)^{2}=\delta\bar{R}'_{j}\delta\bar{R}'_{j_+}+O(\tau )$, in which the subscript $j_+$ denotes a measurement at time $t_j+\tau $ via an apparatus contacting the target from time $t_j$ to $t_j+\tau $. Then we replace
the correlations of $\delta\bar{R}'_{j}$ (not directly observable) with those
of the outputs (directly observable), with an
error vanishing in the WSL. For example, $\tau ^{4}\left\langle \delta\bar{R}'_i\left(\delta\bar{R}'_j\right)^2\delta\bar{R}'_k\right\rangle_S
\approx \tau^4\left\langle\delta\bar{R}'_i\delta\bar{R}'_j\delta\bar{R}'_{j_+}\delta\bar{R}'_k\right\rangle_S\approx  C_{i,j,j_+,k}$ up to an error $\sim O\left(\tau^5\right)$, with the correlation $C_{i,j,j+,k}$ constructed from raw data using Eq.~(\ref{eq:correlation_of_raw_data}).} 
After such reformulation of equal-time correlations in Eq.~(\ref{Eq:inequalities}), we obtain an inequality constructed by experimentally observable quantity $\mathscr{N}(\tau )$ and
$\mathscr{D}(\tau )$ 
\begin{equation}
\mathscr{N}(\tau )+O(\tau ^{3})\le\mathscr{D}(\tau ),\label{Eq:inequalities_final}
\end{equation}
where 
\begin{subequations}
\begin{align}\mathscr{N}(\tau )\equiv & \left|C_{i,j}+C_{j,k}+C_{i,l}-C_{k,l}\right|, \\
\mathscr{D}(\tau )\equiv & \sqrt{\left|\mathscr{A}+2\mathscr{B}\right|},
\end{align}
\label{eq:VBfunction}
\end{subequations}
with
$\mathscr{A}\equiv C_{i,i_+,j,j_+}+C_{j,j_+,k,k_+}+C_{i,i_+,l,l_+}+C_{k,k_+,l,l_+}$ 
and
$\mathscr{B}=C_{i,j,j_+,k}-C_{i,k,l,l_+}+C_{i,i_+,j,l}-C_{j,k,k_+,l}$ 
coming respectively from the square of each term and the cross terms in the right-hand side of
Eq.~(\ref{Eq:inequalities}).
As a result, we find the bound for the ratio $\mathscr{V}(\tau )\equiv\mathscr{N}(\tau )/\mathscr{D}(\tau )$
in the WSL,
\begin{equation}
\lim_{\tau \rightarrow0}\mathscr{V}(\tau )\le 1.\label{eq:legget type}
\end{equation}
The left-hand side of this inequality can be
computed using experimental raw data, for example, the photon
counts or the electric currents recorded by classical devices.
Therefore, it avoids the interpretation loopholes such as the assumption
of the measurement model.  The loophole caused by the NIM assumption
is also closed since the WSL condition $\tau \rightarrow0$ can be verified by checking,
for example, the scaling of magnitude of second-order correlations ($\propto\tau ^{2}$)
and that of fourth-order correlations ($\propto\tau ^{4}$)
{[}Eq.~(\ref{eq:correlations}){]}. {Alternatively, one can perform the measurements sequentially at times $t_n=n\tau$ and combine $p$ consecutive outputs $R_j+R_{j+1}+\cdots+R_{j+p-1}$ to replace the single-shot output $R_j$ in Eq.~(\ref{eq:correlations}). In the WSL, $D_{j}\left(R_j|S\right)=D_{j'}\left(R_{j'}|S\right)+O\left(\tau\right)$ for small $|j-j'|$. Therefore, for small $p$, the $L$-th order correlation should scale as $\left(p\tau\right)^L$ in the WSL, which can be directly verified using experimental raw data.} The scaling of signals with
$\tau \rightarrow 0$ also verifies the independence of the apparatuses
because initial correlations between apparatuses would introduce
a remnant correlation signal even when the interaction is turned off ($\tau \rightarrow 0$). 
The violation of the inequality Eq.~(\ref{eq:legget type})
in the WSL would unambiguously falsify the MR. 

\begin{figure}
\includegraphics[width=0.9\columnwidth]{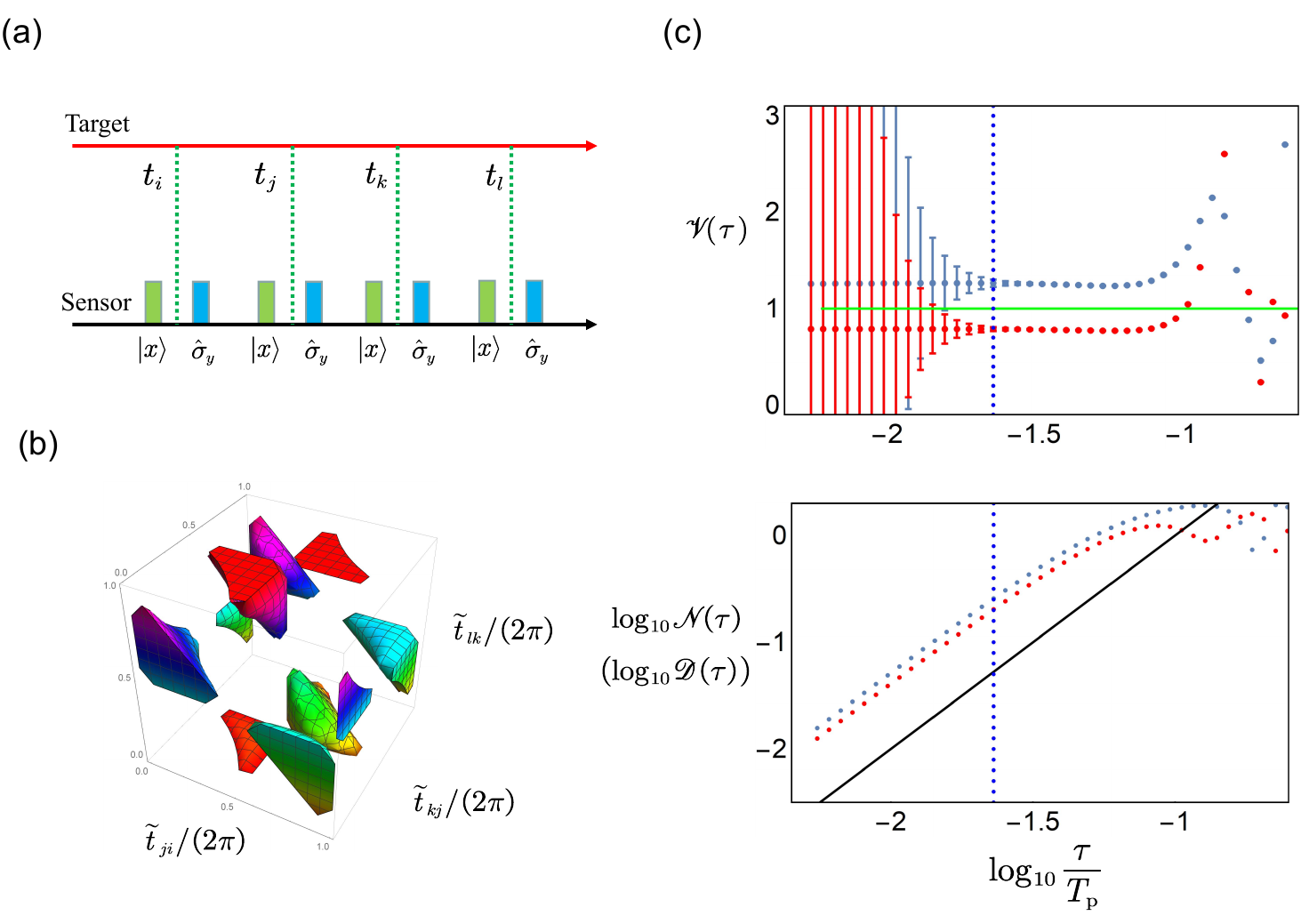}
\caption{\textbf{Loophole-free test of MR}. (a) The measurement
sequence. The green and blue rectangles denote the initialization
and measurement step, respectively. (b) The 3D plot of regions where 
$\lim_{\tau \rightarrow 0}\mathscr{V}(\tau )>1$ for the quantum
spin model as a function of three phase delays ${\tilde t}_{ji}$, ${\tilde t}_{kj}$, and ${\tilde t}_{lk}$.
(c) The upper graph: $\mathscr{V}(\tau )$ as a function of $\log_{10}\left( \tau /T_{\mathrm{p}} \right) $
with ${\tilde t}_{ji}$, ${\tilde t}_{kj}$, and ${\tilde t}_{lk}$. Here $T_{\mathrm{p}}=2\pi /\omega $ is the precession period of the target spin. The blue symbols are for parameters ${\tilde t}_{ji}=2\pi/3,{\tilde t}_{kj}=\pi$, and ${\tilde t}_{lk}=5\pi/3$
and the red symbols are for ${\tilde t}_{ji}=1.2\pi,{\tilde t}_{kj}=0.4\pi$, and ${\tilde t}_{lk}=5\pi/3$. The green line indicates the classical upper bound $\mathscr{V}=1$. The
lower graph: $\log_{10}\mathscr{N}(\tau )$ (blue dots) and $\log_{10}\mathscr{D}(\tau )$ (red dots)
as functions of $\log _{10}\left( \tau /T_{\mathrm{p}} \right)$ for ${\tilde t}_{ji}=2\pi/3,{\tilde t}_{kj}=\pi$, and ${\tilde t}_{lk}=5\pi/3$.
The black line is a guideline with a slope of $2$. The standard deviations are for data collected in $10$~hours under realistic parameters (see main text). For $\tau$ greater than the point marked by the vertical dotted line, the maximal violation (the blue symbols) of the inequality is greater than 5 standard deviations.
The Larmor frequency and target-sensor coupling is set as
$\omega=2\pi$~$\mathrm{\mu}$sec$^{-1}$ and $A_{\bot}=2.2\omega$. \label{fig:Fig2} }
\end{figure}

\textit{Examples.}  We consider
the weak measurement of a single nuclear spin to test
MR. The weak measurement of a target spin-$1/2$ $\hat{\mathbf I}$ can be realized by
coupling it to a sensor spin-$1/2$ $\hat{\mathbf S}$ and then performing a projective measurement of the sensor~\cite{PfenderNC2019,JonasNC2022,WangPRAppl2023}.
Four weak measurements by four independent apparatuses are implemented
consecutively on the target spin at time $t_{1}<t_{2}<t_{3}<t_{4}$.
Each weak measurement includes three steps [see Fig.~\ref{fig:Fig2}(a)]: 1. Prepare the sensor spin to the state
$\vert x\rangle$ ; 2. Entangle the sensor spin and the target
spin via the Hamiltonian $\hat{V}=A_{\perp}\hat{S}_{z}\hat{I}_{x}$
for a duration $\tau $,  where  $A_{\perp}$ is the
coupling strength between the sensor and the target; 3. Measure $\hat{\sigma}_{y}\equiv 2\hat{S}_y/\hbar$
of the sensor and obtain an output $+1$ or $-1$ randomly. The target spin 
at $t_0=0$ is assumed to be fully unpolarized and it evolves under an external field with the Hamiltonian $H_0=\omega\hat{I}_z$, where $\omega$ is the Larmor frequency.  For this model,
the correlations of the outputs can be analytically calculated~\cite{supp}.

The second-order correlation of the outputs is~\cite{supp}
\begin{equation}
C_{i,j}=A_{\perp}^{2}\tau ^{2}\cos\tilde{t}_{ji}+O(\tau ^{3}),\label{eq:S2q}
\end{equation}
where ${\tilde t}_{ji}=\omega(t_j-t_i)$ is the phase delay between two measurements.
The fourth-order correlation 
can be factorized as $C_{i,j,k,l}=C_{i,j}C_{k,l}$~\cite{supp}. Therefore, $\mathscr{V}(\tau )$ {[}Eq.~($\ref{eq:legget type}$){]} in the WSL
becomes 
\begin{equation}
\lim_{\tau \rightarrow0}\mathscr{V}(\tau )=\frac{\left|(\cos{\tilde t}_{ji}+\cos{\tilde t}_{kj})+(\cos({\tilde t}_{ji}+{\tilde t}_{kj}+{\tilde t}_{lk})-\cos{\tilde t}_{lk}\right|}{\sqrt{\left|4+2\left(\sin{\tilde t}_{ji}\sin{\tilde t}_{kj}-\sin{\tilde t}_{kj}\sin{\tilde t}_{lk}\right)\right|}}.\label{eq:limF_quantum}
\end{equation}
Figure~\ref{fig:Fig2}(b) shows the parameter regions where MR is falsified by $\lim_{\tau \rightarrow0}\mathscr{V}(\tau )>1$. 
Because the global
maximum of $\lim_{\tau \rightarrow 0}\mathscr{V}(\tau )$ is
sensitive to the error of the phase delay, 
it is not easy to be reached
experimentally. Instead, we search for the maximum violation of $\lim_{\tau \rightarrow0}\left[\mathscr{N}(\tau )-\mathscr{D}(\tau )\right]/\tau ^{2}<0$, which is 
is reached when ${\tilde t}_{ji}=2\pi/3,{\tilde t}_{kj}=\pi$
and ${\tilde t}_{lk}=5\pi/3$. At this point, $\lim_{\tau \rightarrow0}\mathscr{V}(\tau )=1.25$,  which is larger than the classical bound $\mathscr{V}=1$ {[}green line
in the upper graph of Fig.~\ref{fig:Fig2}(c){]}. Experimentally, we can measure $\mathscr{V}(\tau )$
as a function of $\tau$ for different parameters as shown in the 
upper graph of Fig.~\ref{fig:Fig2} (c)~\cite{supp}.
The WSL can be validated experimentally
by checking the $\tau^2$ scaling of both $\mathscr{N} \left( \tau  \right)$
 and $\mathscr{D} \left( \tau  \right)$ {[}see the lower graph of Fig.~\ref{fig:Fig2}(c){]}. For the parameters ${\tilde t}_{ji}=2\pi/3$, ${\tilde t}_{kj}=\pi$
and ${\tilde t}_{lk}=5\pi/3$, the test $\mathscr{V}(\tau )$ converges to 1.25, in the region where $\mathscr{N}\left( \tau  )\right) \propto \tau ^2$ and $\mathscr{D} \left( \tau  \right) \propto \tau ^2$ {[}lower graph
of Fig.~\ref{fig:Fig2}~(c){]}. The simultaneous occurrence of the $\tau^2$-scaling and $\mathscr{V}>1$ is not allowed by any theory consistent with MR and hence constitutes a loophole-free
rejection of MR.

Now we estimate the effects of shot noises on the test $\mathscr{V}(\tau)$
under realistic conditions. Although the breaking of Eq.~(\ref{eq:legget type})
is illustrated by consecutive measurement sequences, we can also realize
it by sequential weak measurements since the signals are the same at the WSL~\cite{PfenderNC2019,JonasNC2022}. Taking sequential measurement 
in an nitrogen-vacancy center in diamond~\cite{PfenderNC2019,JonasNC2022,RobledoNature2011} as
an example, the uncertainty of the fourth-order signal is estimated to be $\left(\tau/T\right)^{1/2}/\left(\sqrt{\chi_{\rm ph} \tau}\right)^{4}$,
where $\chi_{\rm ph}$ is the photon count rate and $T$ is total data
acquisition time. For realistic parameters such as $T_{p}=2\pi/\omega\approx 1~\mu$sec, 
$\tau/T_{p}=0.023$,
photon count rate $\chi_{\rm ph}\approx 3\times 10^6$~sec$^{-1}$~\cite{WangSA2024,RobledoNature2011} , $T\approx 10 $~hours and the coupling in Fig. \ref{fig:Fig2}, one
can achieve an uncertainty of the fourth-order signal $\sim 2.4\times 10^{-3}$,
which is much smaller than the magnitude of the signal $\mathscr{D}(\tau)^2\sim 4\times 10^{-2}$.
Based on the uncertainties of the signals, the uncertainty of $\mathscr{V}(\tau)$
is shown in the upper graph of Fig.~\ref{fig:Fig2}(c). We can see that it is possible to detect a large scaling region $0.018\le\tau/T_p\le 0.056$
with a $10^{2}$-fold variation of the magnitude of the fourth-order signal.
At the point $\tau= 0.023 T_p$, e.g., the inequality in Eq.~(\ref{eq:legget type}) can be violated by 6.5 standard deviations, i.e., with a statistical confidence $>1-4\times 10^{-11}$.

It should be emphasized that the second-order correlation {[}Eq.~(\ref{eq:S2q}){]}
of the raw data can never break the original LG inequality~\cite{LeggettPRL1985}
since the correlation approaches zero as $\tau \rightarrow0$.
As a result, an apparatus-dependent constant $A_{\perp}^{2}\tau ^{2}$
should be removed to break the LG inequality~\cite{PalaciosNP2010}.
However, it introduces he following loopholes. First, a binary measurement model is assumed. Second, the characterization of the constant $A_{\perp}^{2}\tau ^{2}$
requires the concept of quantum mechanism~\cite{PalaciosNP2010}. Third, more
seriously, the NIM loophole cannot be excluded~\cite{PalaciosNP2010}.
In contrast, the bound in Eq.~(\ref{eq:legget type}) is a general
bound following the MR principle without additional assumptions and therefore
can provide a loophole-free test of the MR principle. 

For comparison, we also give an example to show that a
classical system complies with the MR principle and can never violate the
inequality Eq.~(\ref{eq:legget type}). We consider a classical version
of the quantum spin model with all the quantities replaced
by their classical correspondence. The classical model can be described by the Liouville density, so it satisfies the MR principle.
Following the same steps as in the quantum case, we obtain~\cite{supp}
\begin{equation}
\lim_{\tau \rightarrow0}\mathscr{V}(\tau )=\frac{\sqrt{15}}{6}\frac{\left|\cos{\tilde t}_{ji}+\cos{\tilde t}_{kj}+\cos {\tilde t}_{li}-\cos{\tilde t}_{lk}\right|}{\sqrt{\left|2+a+b\right|}},\label{eq:boundc}
\end{equation}
with $a\equiv\cos({\tilde t}_{ji}+{\tilde t}_{kj})\cos({\tilde t}_{ji}+{\tilde t}_{lk})\cos({\tilde t}_{kj}+{\tilde t}_{lk})$
and $b\equiv\left( \sin {\tilde t}_{ji}-\sin {\tilde t}_{lk} \right) \times \left( \sin {\tilde t}_{kj}-\sin {\tilde t}_{li}\right) $.
The maximum of $\lim_{\tau \rightarrow0}\mathscr{V}(\tau )$
is $0.913$, 
which is consistent with the prediction
that the classical system can never break the inequality Eq.~($\ref{eq:legget type}$).
Interestingly, the behavior of the second-order signal of this classical model is
the same as that of the quantum spin model, 
which suggests that the high-order correlations are essential for the loophole-free test of MR.

\textit{Conclusion}. We have
proposed an inequality based on high-order correlations for loophole-free test of MR. Simultaneous violation of this inequality and witness
of expected scaling of correlation signals of different orders would unambiguously falsify
MR. We illustrate the violation of this inequality by a quantum model.
 Experimental realization of this protocol is feasible
in quantum systems such as superconductor qubits~\cite{PalaciosNP2010},
cold atoms~\cite{YangNature2020}, color center spins in diamond~\cite{PfenderNC2019,JonasNC2022},
photonic qubits~\cite{GogginPNAS2011,SuzukiNJP2012} and spins in quantum dots~\cite{BechtoldPRL2016}.
The method proposed here can be generalized to obtain a class of inequalities
between multi-point correlations~\cite{EmaryRPP2013} and can also be used
to develop new methods to characterize macroscopic quantum coherence~\cite{FroewisRMP2018}. 

\textit{Acknowledge}. P. W. is supported by the Talents
Introduction Foundation of Beijing Normal University with Grant No.
310432106 and partially supported by the Hong Kong RGC General Research Fund (14300119). R. B. L. was supported by the Hong Kong RGC Senior Research Fellow Scheme and the New Cornerstone Science Foundation. J. W. acknowledges financial support from EU Project AMADEUS and QIA, DFG (GRK2642), and DFG Research group FOR 2724 as well as the BMBF via the projects SPINNING and QRX. H. L acknowledges financial support from the National Natural Science Foundation of China (62276171) and CCF-Baidu Open Fund (Grant No. OF2022028), and Swiftlet Fund Fintech funding.

\bibliographystyle{apsrev4-1}
%


\end{document}


\title{Supplementary Materials for \textquotedblleft Loophole-free test of macroscopic realism via high-order correlations of measurement \textquotedblright{}}
\today

\author{Ping Wang}
\email{wpking@bnu.edu.cn}
\affiliation{College of Education for the Future, Beijing Normal University, Zhuhai 519087, China}
\affiliation{Department of Physics, The Chinese University of Hong Kong, Shatin, New Territories, Hong Kong, China}
\affiliation{Centre for Quantum Coherence,
The Chinese University of Hong Kong, Shatin, New Territories, Hong Kong, China}
\affiliation{The Hong Kong Institute of Quantum Information Science and Technology,
The Chinese University of Hong Kong, Shatin, New Territories, Hong Kong, China}

\author{Chong Chen}
\affiliation{Department of Physics, The Chinese University of Hong Kong, Shatin, New Territories, Hong Kong, China}
\affiliation{Centre for Quantum Coherence,
The Chinese University of Hong Kong, Shatin, New Territories, Hong Kong, China}
\affiliation{The Hong Kong Institute of Quantum Information Science and Technology,
The Chinese University of Hong Kong, Shatin, New Territories, Hong Kong, China}

\author{Hao Liao}
\affiliation{National Engineering Laboratory on Big Data System Computing Technology, Guangdong Province Engineering Center of
China-made High Performance Data Computing System, College of Computer Science and Software Engineering, Shenzhen
University, Shenzhen 518060, China}

\author{Vadim V. Vorobyov}
\affiliation{3rd Institute of Physics, Research Center SCoPE and IQST, University of Stuttgart, 70569 Stuttgart, Germany}

\author{ J\"org Wrachtrup}
\affiliation{3rd Institute of Physics, Research Center SCoPE and IQST, University of Stuttgart, 70569 Stuttgart, Germany}

\author{Ren-Bao Liu}
\email{rbliu@cuhk.edu.hk}
\affiliation{Department of Physics, The Chinese University of Hong Kong, Shatin, New Territories, Hong Kong, China}
\affiliation{Centre for Quantum Coherence,
The Chinese University of Hong Kong, Shatin, New Territories, Hong Kong, China}
\affiliation{The Hong Kong Institute of Quantum Information Science and Technology,
The Chinese University of Hong Kong, Shatin, New Territories, Hong Kong, China}
\affiliation{New Cornerstone Science Laboratory,
The Chinese University of Hong Kong, Shatin, New Territories, Hong Kong, China}
\maketitle


\subsection{Conditional probabilities of outputs in the weak signal limit (WSL)}

Here we prove Eq.~(2) in main text. We assume the $j$-th apparatus contacts with the target system from $t_{j}-\tau_{j}$ to
$t_{j}$ and the output $R_{j}$ is recorded at $t_{j}$. All the
measurement times $\tau_{j}$'s are assumed to be short (the WSL condition).
If the $n$-th apparatus has no interaction with the target system ($\tau_{n}=0$),
its output would be independent of the state of the target system and the outputs of the previous measurements, that is, $P_{n}\left(R_{n}|{\mathbf R}^{(n-1)},S\right)
\rightarrow P_n(R_n)$ at the limit $\tau_n\rightarrow 0$.
Therefore, by Taylor expansion with respect to $\tau_n$,
the conditional probability 
\[
P_{n}\left(R_{n}|{\mathbf R}^{(n-1)},S\right)
=P_{n}\left(R_{1}\right)+\tau_{n}K_{n}\left(R_{n}|{\mathbf R}^{(n-1)},S\right)+O\left(\tau_{n}^{2}\right),
\]
where $K_{n}\left(R_{n}|{\mathbf R}^{(n-1)},S\right)\equiv\left. \partial_{\tau_n}P_{n}\left(R_{n}|{\mathbf R}^{(n-1)},S\right)\right|_{\tau_n=0} $ denotes the correlation between
the $n$-th apparatus and the target system and the previous measurements due to the coupling at time $t_n$. Note that $K_n$ is a function of $\tau_1,\tau_2,\ldots,\tau_{n-1}$. When $\tau_1=\tau_2=\cdots=\tau_{n-1}=0$ (i.e., no measurement is performed before $t_n$), the conditional probability $P_{n}\left(R_{n}|{\mathbf R}^{(n-1)},S\right)$ should be independent of the earlier outputs 
${\mathbf R}^{(n-1)}$, so $K_{n}\left(R_{n}|{\mathbf R}^{(n-1)},S\right)\rightarrow D_n\left(R_n|S\right)$. Again, by Taylor expansion with respect to $(\tau_1,\tau_2,\ldots,\tau_{n-1})$, 
$$
K_{n}\left(R_{n}|{\mathbf R}^{(n-1)},S\right)=D\left(R_n|S\right)+O\left(\tau_1\right)+O\left(\tau_2\right)+\cdots+O\left(\tau_{n-1}\right).
$$
Letting $\tau_1=\tau_2=\cdots =\tau_n=\tau$, we get
\[
P_{n}\left(R_{n}|{\mathbf R}^{(n-1)},S\right)
=P_{n}\left(R_{1}\right)+\tau D_{n}\left(R_{n}|S\right)+O\left(\tau^{2}\right).
\]
Thus, Eq.~(2) in main text is proved.

\subsection{Correlation signals of the quantum spin model}

\subsubsection{General formalism}

The first-order correlations (expectation values) vanish for the quantum spin
model for the measurement setup in main text\cite{PfenderNC2019}.

The second order correlations defined in main text are\cite{PfenderNC2019,JonasNC2022}
\[
\begin{aligned}C_{i,j}= & \mathrm{Tr}\mathcal{K}\mathcal{U}_{t_{ji}-\tau}\mathcal{K}\mathcal{U}_{t_{i}-\tau}\hat{\rho}_{\mathrm{th}},\\
C_{i,k}= & \mathrm{Tr}\mathcal{K}\mathcal{U}_{t_{kj}-\tau}\mathcal{M}\mathcal{U}_{t_{ji}-\tau}\mathcal{K}\mathcal{U}_{t_{i}-\tau}\hat{\rho}_{\mathrm{th}},\\
C_{i,l}= & \mathrm{Tr}\mathcal{K}\mathcal{U}_{t_{lk}-\tau}\mathcal{M}\mathcal{U}_{t_{kj}-\tau}\mathcal{M}\mathcal{U}_{t_{ji}-\tau}\mathcal{K}\mathcal{U}_{t_{i}-\tau}\hat{\rho}_{\mathrm{th}},\\
C_{j,k}= & \mathrm{Tr}\mathcal{K}\mathcal{U}_{t_{kj}-\tau}\mathcal{K}\mathcal{U}_{t_{ji}-\tau}\mathcal{M}\mathcal{U}_{t_{i}-\tau}\hat{\rho}_{\mathrm{th}},\\
C_{j,l}= & \mathrm{Tr}\mathcal{K}\mathcal{U}_{t_{lk}-\tau}\mathcal{M}\mathcal{U}_{t_{kj}-\tau}\mathcal{K}\mathcal{U}_{t_{ji}-\tau}\mathcal{M}\mathcal{U}_{t_{i}-\tau}\hat{\rho}_{\mathrm{th}},\\
C_{k,l}= & \mathrm{Tr}\mathcal{K}\mathcal{U}_{t_{lk}-\tau}\mathcal{K}\mathcal{U}_{t_{kj}-\tau}\mathcal{M}\mathcal{U}_{t_{ji}-\tau}\mathcal{M}\mathcal{U}_{t_{i}-\tau}\hat{\rho}_{\mathrm{th}},
\end{aligned}
\]
where $\hat{\rho}_{\mathrm{th}}=1/2$ denotes the initial states of
the spin, $\mathcal{U}_{t}\hat{\rho}\equiv e^{-i\omega t\hat{I}_{z}}\hat{\rho} e^{i\omega t\hat{I}_{z}}$
accounts for the free evolution between two measurements, 
the super-operator $\mathcal{K}$ accounts for the measurement, and
$\mathcal{M}$ denotes the effect of measurement when the outputs are
discarded (i.e., the measurement-induced decoherence). $\mathcal{K}$ and $\mathcal{M}$ can be formulated as\cite{PfenderNC2019,JonasNC2022}
\begin{subequations}
\begin{align}\mathcal{K}\hat{\rho}= & \frac{1}{2i}\left[\hat{U}_{-}\hat{\rho}\hat{U}_{+}^{\dagger}-h.c\right],\\
\mathcal{M}\hat{\rho}= & \frac{1}{2}\left[\hat{U}_{+}\hat{\rho}U_{+}^{\dagger}+U_{-}\hat{\rho}U_{-}^{\dagger}\right],
\end{align}
\label{eq:K and M}
\end{subequations}
where the path-dependent evolution operator $\hat{U}_{\pm}$ in the
interaction process is
\begin{equation}
\hat{U}_{\pm}=e^{-i\hat{H}_{\pm}\tau}\equiv e^{-i\phi\mathbf{n}_{\pm}\cdot\hat{\mathbf{I}}},\label{eq:Upm}
\end{equation}
with $\hat{H}_{\pm}=\omega\hat{I}_{z}\pm A_{\perp}\hat{I}_{x}$ being
the Hamiltonian of the target spin conditioned on the sensor spin state. In the equation above
$\mathbf{n}_{\pm}\equiv\cos\theta\mathbf{e}_{z}\pm\sin\theta\mathbf{e}_{x}$ (with $\tan\theta=A_{\perp}/\omega$)
and $\phi=\sqrt{A_{\perp}^{2}+\omega^{2}}\tau$.

In the model considered, $\hat{\rho}_{\mathrm{th}}=1/2$ is the steady
state of both $\mathcal{M}$ and $\mathcal{U}$ (that is, $\mathcal{M}\hat{\rho}_{\mathrm{th}}=\hat{\rho}_{\mathrm{th}}$
and $\mathcal{U}\hat{\rho}_{\mathrm{th}}=\hat{\rho}_{\mathrm{th}}$).
The second-order correlations can be reformulated as 
\begin{subequations}
\begin{align}C_{i,j}= & \mathrm{Tr}\tilde{\mathcal{K}}\mathcal{U}_{t_{ji}}\tilde{\mathcal{K}}\hat{\rho}_{\mathrm{th}},\\
C_{i,k}= & \mathrm{Tr}\tilde{\mathcal{K}}\mathcal{U}_{t_{kj}}\tilde{\mathcal{M}}\mathcal{U}_{t_{ji}}\tilde{\mathcal{K}}\hat{\rho}_{\mathrm{th}},\\
C_{i,l}= & \mathrm{Tr}\tilde{\mathcal{K}}\mathcal{U}_{t_{lk}}\tilde{\mathcal{M}}\mathcal{U}_{t_{kj}}\tilde{\mathcal{M}}\mathcal{U}_{t_{ji}}\tilde{\mathcal{K}}\hat{\rho}_{\mathrm{th}},\\
C_{j,k}= & \mathrm{Tr}\tilde{\mathcal{K}}\mathcal{U}_{t_{kj}}\tilde{\mathcal{K}}\hat{\rho}_{\mathrm{th}}\equiv C_{i,j}\vert_{j\rightarrow k,i\rightarrow j},\\
C_{j,l}= & \mathrm{Tr}\tilde{\mathcal{K}}\mathcal{U}_{t_{lk}}\tilde{\mathcal{M}}\mathcal{U}_{t_{kj}}\tilde{\mathcal{K}}\hat{\rho}_{\mathrm{th}}\equiv C_{i,k}\vert_{k\rightarrow l,i\rightarrow j},\\
C_{k,l}= & \mathrm{Tr}\tilde{\mathcal{K}}\mathcal{U}_{t_{lk}}\tilde{\mathcal{K}}\hat{\rho}_{\mathrm{th}}\equiv C_{i,j}\vert_{j\rightarrow l,i\rightarrow k},
\end{align}
\label{eq:corr2}
\end{subequations}
where $\tilde{\mathcal{K}}\equiv\mathcal{U}_{-\tau}\mathcal{K}$ and $\tilde{\mathcal{M}}\equiv\mathcal{U}_{-\tau}\mathcal{M}$.
The combination of $\mathcal{U}^{-\tau}$ and $\mathcal{K},\mathcal{M}$
is used to compensate the free evolution of target spin of $\mathcal{K},\mathcal{M}$,
which is essentially the same to define super-operator in the interaction
picture.

Similarly, the fourth-order correlations can be formulated
as 
\begin{align}
C_{i,j,k,l} &=\mathrm{Tr}\mathcal{K}\mathcal{U}_{t_{lk}-\tau}\mathcal{K}\mathcal{U}_{t_{kj}-\tau}\mathcal{K}\mathcal{U}_{t_{ji}-\tau}\mathcal{K}\mathcal{U}_{t_{i}-\tau}\hat{\rho}_{\mathrm{th}}
\nonumber\\
&=\mathrm{Tr}\tilde{\mathcal{K}}\mathcal{U}_{t_{lk}}\tilde{\mathcal{K}}\mathcal{U}_{t_{kj}}\tilde{\mathcal{K}}\mathcal{U}_{t_{ji}}\tilde{\mathcal{K}}\hat{\rho}_{\mathrm{th}}.\label{eq:corr4}
\end{align}

We can calculate the correlation signals and $\mathscr{V}(\tau)$ using
Eq.~(\ref{eq:corr2}) and Eq.~(\ref{eq:corr4}).

\subsubsection{Correlation signals in the WSL}

When $\tau\rightarrow0$, 
the evolution operator
{[}Eq.~(\ref{eq:Upm}){]} reduces to 
\begin{equation}
\hat{U}_{\pm}\approx e^{-i\omega\tau\hat{I}_{z}}\left(e^{i\omega\tau\hat{I}_{z}}e^{-i\phi\mathbf{n}_{\pm}\cdot\hat{\mathbf{I}}}\right)=e^{-i\omega\tau\hat{I}_{z}}\left(1\pm iA_{\perp}\tau\mathbf{e}_{x}\cdot\hat{\mathbf{I}}\right)+O(\tau^{2}). \label{eq:Upm_approx}
\end{equation}
As a result, we obtain
\begin{subequations}
\begin{align}\mathcal{K}\hat{\rho}= & A_{\perp}\tau\mathcal{U}_{\tau}\left\{ \mathbf{e}_{x}\cdot\hat{\mathbf{I}},\hat{\rho}\right\} +O\left(\tau^2\right),\\
\mathcal{M}\hat{\rho}= & \mathcal{U}_{\tau}\hat{\rho}+O(\tau),
\end{align}
\label{eq:K and M approx}
\end{subequations}
so
\begin{subequations}
\begin{align}\tilde{\mathcal{K}}\hat{\rho}= & A_{\perp}\tau\left\{ \mathbf{e}_{x}\cdot\hat{\mathbf{I}},\hat{\rho}\right\}+O\left(\tau^2\right), \\
\tilde{\mathcal{M}}\hat{\rho}= & \hat{\rho}+O(\tau).
\end{align}
\label{eq:K and M tilde approx}
\end{subequations}
Using this formula, we find
\begin{subequations}
\begin{align}
C_{i,j}= & \mathrm{Tr}\tilde{\mathcal{K}}\mathcal{U}_{t_{ji}}\tilde{\mathcal{K}}\hat{\rho}_{\mathrm{th}}=A_{\perp}^{2}\tau^{2}\cos\omega t_{ji}+O(\tau^{3}),\\
C_{i,k}= & \mathrm{Tr}\tilde{\mathcal{K}}\mathcal{U}_{t_{ki}}\tilde{\mathcal{K}}\hat{\rho}_{\mathrm{th}}+O(\tau^{3})=A_{\perp}^{2}\tau^{2}\cos\omega t_{ki}+O(\tau^{3}),\\
C_{i,l}= & \mathrm{Tr}\tilde{\mathcal{K}}\mathcal{U}_{t_{li}}\tilde{\mathcal{K}}\hat{\rho}_{\mathrm{th}}+O(\tau^{3})=A_{\perp}^{2}\tau^{2}\cos\omega t_{li}+O(\tau^{3}),\\
C_{j,k}= & \mathrm{Tr}\tilde{\mathcal{K}}\mathcal{U}_{t_{kj}}\tilde{\mathcal{K}}\hat{\rho}_{\mathrm{th}}+O(\tau^{3})\equiv A_{\perp}^{2}\tau^{2}\cos\omega t_{li}+O(\tau^{3}),\\
C_{j,l}= & \mathrm{Tr}\tilde{\mathcal{K}}\mathcal{U}_{t_{lj}}\tilde{\mathcal{K}}\hat{\rho}_{\mathrm{th}}+O(\tau^{3})=A_{\perp}^{2}\tau^{2}\cos\omega t_{lj}+O(\tau^{3}),\\
C_{k,l}= & \mathrm{Tr}\tilde{\mathcal{K}}\mathcal{U}_{t_{lk}}\tilde{\mathcal{K}}\hat{\rho}_{\mathrm{th}}+O(\tau^{3})=A_{\perp}^{2}\tau^{2}\cos\omega t_{lk}+O(\tau^{3}).
\end{align}
\label{eq:C2approx}
\end{subequations}
Similarly, the fourth-order correlation in the leading order of $\tau$ becomes
\begin{align}
C_{i,j,k,l}\approx & \mathrm{Tr}\tilde{\mathcal{K}}\mathcal{U}_{t_{lk}}\tilde{\mathcal{K}}\mathcal{U}_{t_{kj}}\tilde{\mathcal{K}}\mathcal{U}_{t_{ji}}\tilde{\mathcal{K}}\hat{\rho}_{\mathrm{th}}\nonumber\\
= & \mathrm{Tr}\tilde{\mathcal{K}}\mathcal{U}_{t_{lk}}\tilde{\mathcal{K}}\mathcal{U}_{t_{kj}}\left\{ \hat{\rho}_{\mathrm{th}}\mathrm{Tr}\left[\tilde{\mathcal{K}}\mathcal{U}_{t_{ji}}\tilde{\mathcal{K}}\hat{\rho}_{\mathrm{th}}\right]\right\} \nonumber \\
=&C_{kl}C_{ji}.
\label{eq:corr4-approx}
\end{align}
Consequently,
we have 
\begin{equation}
C_{i,j,k,l}=A_{\perp}^{4}\tau^{4}\cos\omega t_{lk}\cos\omega t_{ji}+O(\tau^{5}).\label{eq:corr4-approx2}
\end{equation}
Using Eqs.~(\ref{eq:C2approx}) and (\ref{eq:corr4-approx2}),
we obtain $\mathscr{V}(\tau)$ under in the WSL.

\begin{figure}
\includegraphics[width=0.6\columnwidth]{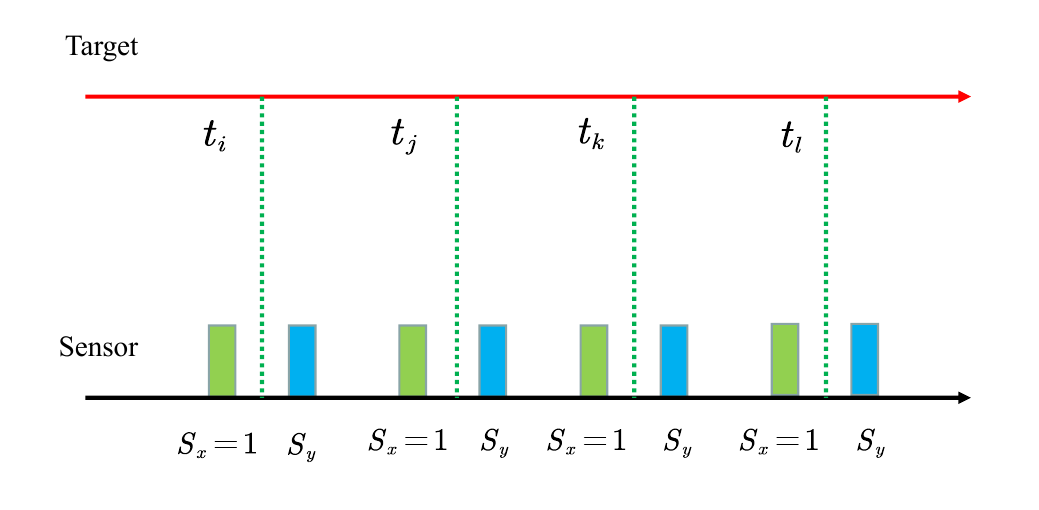}
\caption{\textbf{Measurement sequence for the classical spin model.} In each
measurement, the apparatus begins with a given initial state ($S_{x}=1$,
indicated by a green box), then interacts with the target for a short duration
$\tau$, and finally the observable $S_{y}$ of the apparatus is recorded (indicated by a blue box). Based on these records, the correlation of
the outputs is calculated. \label{fig:FigS1}}
\end{figure}

\subsection{Correlation signals of the classical spin model}

\subsubsection{Dynamics of the classical spin model}

For the classical spin model, we assume that the system has the
Hamiltonian with the same form as the quantum model
\begin{equation}
H=\omega I_{z}+A_{\perp}S_{z}I_{x},\label{eq:Hami_classical}
\end{equation}
but with $\mathbf{I}=\mathbf{r}_{\mathrm{I}}\times\mathbf{p}_{\mathrm{I}}$
and $\mathbf{S}=\mathbf{r}_{\mathrm{S}}\times\mathbf{p}_{\mathrm{S}}$
being classical angular momenta of the target and the apparatus, respectively, where
$\{\mathbf{r}_{\mathrm{I}},\mathbf{p}_{\mathrm{I}}\}$ and $\{\mathbf{r}_{\mathrm{S}},\mathbf{p}_{\mathrm{S}}\}$
denote the canonical positions and momenta of the target and the apparatus,
respectively. 
It is easy to shown that $I^2$ and $S^2$
are conserved quantities. Without loss of generality, we set $S^2=I^2=1$. 

The measurement sequence is shown in Fig.~\ref{fig:FigS1}.
The initial state of the apparatus before each measurement is set as $S_{x}=1$, $S_{y}=S_z=0$.
We assume that the initial state of the target distributes homogeneously
in the space of the angular momentum, which is the classical correspondence
for an unpolarized quantum spin. Mathematically, if we define $I_{x}=I\sin\theta\cos\varphi$,
$I_{y}=I\sin\theta\sin\varphi$ and $I_{z}=I\cos\theta$, the distribution of $\theta,\varphi$ is $P(\theta,\varphi)=1/(4\pi)$
with
$
\int_{0}^{\pi}\sin\theta d\theta\int_{0}^{2\pi}d\varphi P(\theta,\varphi)=1.
$

The dynamics of the angular momenta is governed by the Hamiltonian equations
\begin{subequations}
\begin{align}
\frac{d\mathbf{I}}{dt} & =\left(\omega\mathbf{e}_{z}+AS_{z}\mathbf{e}_{x}\right)\times\mathbf{I}=\overline{\boldsymbol{\omega}}\times\mathbf{I},\label{eq:cd1}
\\
\frac{d\mathbf{S}}{dt} & =\left\{ H,\mathbf{S}\right\} _{p}=AI_{x}\mathbf{e}_{z}\times\mathbf{S},\label{eq:cd2}
\end{align}
\end{subequations}
where $\overline{\boldsymbol{\omega}}\equiv\omega\mathbf{e}_{z}+A_{\perp}S_{z}\mathbf{e}_{x}$
is  the effective angular frequency of the target.
With the initial condition of the apparatus ${S}_{x}=1$ and $S_y=S_{z}=0$,
we find $S_{z}(t)=0$ from Eq.~(\ref{eq:cd2}). Then, from Eq.~(\ref{eq:cd1}),
the dynamics of $\mathbf{I}$ can be solved as
\begin{subequations}
\begin{align}I_{x}(t)= & \cos(\omega t) I_{x}(0)-\sin (\omega t) I_{y}(0),\\
I_{y}(t)= & \sin (\omega t) I_{x}(0)+\cos (\omega t) I_{y}(0),\\
I_{z}(t)= & I_{z}(0).
\end{align}
\label{eq:cd3}
\end{subequations}
Here $I_{x}(0)=\sin\theta\cos\varphi$, $I_{x}(0)=\sin\theta\sin\varphi$, and $I_{z}(0)=\cos\theta$.
Using Eq.~(\ref{eq:cd3}), the $x$ and $y$ components of the apparatus can
also solved from Eq.~(\ref{eq:cd2}). The dynamics of apparatus is
\[
\begin{aligned}S_{x}(t)= & \cos\Phi(t),\\
S_{y}(t)= & \sin\Phi(t),\\
S_{z}(t)= & 0,
\end{aligned}
\]
 where 
\[
\Phi(t)\equiv A_{\perp}\int_{t-\tau}^{t}I_{x}{(u)}du=\left(\frac{2A_{\perp}}{\omega}\sin\theta\sin\frac{\omega\tau}{2}\right)\cos\left(\omega t+\varphi-\frac{\omega\tau}{2}\right).
\]
 Hence the effect of the target is similar to an AC magnetic
field\cite{JonasNC2022}.

\subsubsection{Correlation signals in the WSL}

The observable is chosen to be $S_{y}(t)=\sin\Phi(t)$ of the apparatus.
In the WSL ($\tau\rightarrow0$), $\Phi(t)$ can be approximated as
\[
\Phi(t)\approx A_{\perp}\int_{t-\tau}^{t}I_{x}(u)du=A_{\perp}\tau\sin\theta\cos\left(\omega t+\varphi\right),
\]
up to an error $\sim O\left(\tau^2\right)$. Obviously, the expectation value $\left\langle S_{y}(t)\right\rangle=0$.
The second-order correlation signal is 
\[
\begin{aligned}C_{i,j}= & \left\langle\Phi(t_{j})\Phi(t_{i})\right\rangle+O(\tau^{3})\\
= & A_{\perp}^{2}\tau^{2}\left\langle \sin^{2}\theta\right\rangle _{\theta}\left\langle\cos\left(\omega t_{j}+\varphi\right)\cos\left(\omega t_{i}+\varphi\right)\right\rangle_{\varphi}+O(\tau^{3})\\
= & \frac{2A_{\perp}^{2}\tau^{2}}{3}\left\langle\cos\left(\omega t_{j}+\varphi\right)\cos\left(\omega t_{i}+\varphi\right)\right\rangle_{\varphi}+O(\tau^{3}),
\end{aligned}
\]
 where $\left\langle\right\rangle_{\theta/\varphi}$ denotes the ensemble averaging
over $\theta$ or $\varphi$. The averaging over $\varphi$
leads to 
\begin{equation}
\left\langle\cos\left(\omega u_{1}+\varphi\right)\cos\left(\omega u_{2}+\varphi\right)\right\rangle_{\varphi}=  \sum_{s_{1},s_{2}=\pm1}\frac{1}{4}e^{i\omega\left(s_{1}t_{1}+s_{2}t_{2}\right)}\left\langle e^{i\left(s_{1}+s_{2}\right)\varphi}\right\rangle_{\varphi}
=  \frac{1}{2}\cos\omega(u_{2}-u_{1}),
\label{eq:COS2}
\end{equation}
for any $u_{1},u_{2}$. Therefore, 
\begin{equation}
C_{i,j}=\frac{A_{\perp}^{2}\tau^{2}}{3}\cos\left[\omega t_{ji}\right]+O(\tau^{3}).\label{eq:S2approx_classical}
\end{equation}
Following the same steps, the fourth-order correlation signal is derived as
\[
C_{i,j,k,l}=\frac{8A_{\perp}^{4}\tau^{4}}{15}\left\langle\cos\left(\omega t_{i}+\varphi\right)\cos\left(\omega t_{j}+\varphi\right)\cos\left(\omega t_{k}+\varphi\right)\cos\left(\omega t_{l}+\varphi\right)\right\rangle_{\varphi}+O(\tau^{5}).
\]
Using the identity in Supplementary Information of Ref.~[\onlinecite{JonasNC2022}] 
\[
\begin{aligned}\left\langle\prod_{i=1}^{4}\cos\left(\omega u_{i}+\varphi\right)\right\rangle_{\varphi}= & \frac{1}{2}\left\langle\cos\left(\omega u_{1}+\varphi\right)\cos\left(\omega u_{2}+\varphi\right)\right\rangle_{\varphi}\left\langle\cos\left(\omega u_{3}+\varphi\right)\cos\left(\omega u_{4}+\varphi\right)\right\rangle_{\varphi}\\
+ & \frac{1}{2}\left\langle\cos\left(\omega u_{1}+\varphi\right)\cos\left(\omega u_{3}+\varphi\right)\right\rangle_{\varphi}\left\langle\cos\left(\omega u_{2}+\varphi\right)\cos\left(\omega u_{4}+\varphi\right)\right\rangle_{\varphi}\\
+ & \frac{1}{2}\left\langle\cos\left(\omega u_{1}+\varphi\right)\cos\left(\omega u_{4}+\varphi\right)\right\rangle_{\varphi}\left\langle\cos\left(\omega u_{2}+\varphi\right)\cos\left(\omega u_{3}+\varphi\right)\right\rangle_{\varphi},
\end{aligned}
\]
and the result of Eq.~(\ref{eq:COS2}), we obtain the fourth-order correlation signal in the WSL
\begin{equation}
C_{i,j,k,l}=\frac{A_{\bot}^{4}\tau^{4}}{15}\left(\cos\tilde{t}_{ji}\cos\tilde{t}_{lk}+\cos\tilde{t}_{ki}\cos\tilde{t}_{lj}+\cos\tilde{t}_{li}\cos\tilde{t}_{kj}\right)+O(\tau^{5}), \label{eq:S4approx_classical}
\end{equation}
in which $\tilde{t}_{ji}\equiv \omega t_{ji}$.

\subsubsection{$\mathscr{V}(\tau)$ for the classical model in the WSL}

From Eqs.~(\ref{eq:S2approx_classical}) and (\ref{eq:S4approx_classical}),
the numerator $\mathscr{N}(\tau)$ becomes
\[
\begin{aligned}\mathscr{N}(\tau)= & \left|C_{i,j}+C_{j,k}+C_{i,l}-C_{k,l}\right|\\
= & \frac{A_{\perp}^{2}\tau^{2}}{3}\left|\cos\varphi_{1}+\cos\varphi_{2}+\cos(\varphi_{1}+\varphi_{2}+\varphi_{3})-\cos\varphi_{3}\right|+O(\tau^{3}),
\end{aligned}
\]
where $\varphi_{1}=\tilde{t}_{ji}$, $\varphi_{2}=\tilde{t}_{kj}$
and $\varphi_{3}=\tilde{t}_{lk}$.

The fourth-order correlations appearing in $\mathscr{D}(\tau)$ are
\[
\begin{aligned}C_{i,i_{+},j,j_{+}}= & \frac{A_{\perp}^{4}\tau^{4}}{15}\left[2+\cos(2\varphi_{1})\right]+O(\tau^{5}),\\
C_{i,i_{+},l,l_{+}}= & \frac{A_{\perp}^{4}\tau^{4}}{15}\left[2+\cos[2(\varphi_{1}+\varphi_{2}+\varphi_{3})]\right]+O(\tau^{5}),\\
C_{j,j_{+},k,k_{+}}= & \frac{A_{\perp}^{4}\tau^{4}}{15}\left[2+\cos2\varphi_{2}\right]+O(\tau^{5}),\\
C_{k,k_{+},l,l_{+}}= & \frac{A_{\perp}^{4}\tau^{4}}{15}\left[2+\cos2\varphi_{3}\right]+O(\tau^{5}),
\end{aligned}
\]
and 
\[
\begin{aligned}C_{i,j,j_{+},k}= & \frac{A_{\perp}^{4}\tau^{4}}{15}\left[2\cos\varphi_{1}\cos\varphi_{2}+\cos(\varphi_{1}+\varphi_{2})\right]+O(\tau^{5}), \\
C_{i,k,l,l_{+}}= & \frac{A_{\perp}^{4}\tau^{4}}{15}\left[2\cos(\varphi_{1}+\varphi_{2})+\cos(\varphi_{1}+\varphi_{2}+2\varphi_{3})\right]+O(\tau^{5}), \\
C_{i,i_{+},j,l}= & \frac{A_{\perp}^{4}\tau^{4}}{15}\left[2\cos(\varphi_{2}+\varphi_{3})+\cos(2\varphi_{1}+\varphi_{2}+\varphi_{3})\right]+O(\tau^{5}), \\
C_{j,k,k_{+},l}= & \frac{A_{\perp}^{4}\tau^{4}}{15}\left[2\cos\varphi_{2}\cos\varphi_{3}+\cos(\varphi_{2}+\varphi_{3})\right]+O(\tau^{5}).
\end{aligned}
\]
From these results, we calculate $\mathscr{D}(\tau)$
through the formula
\[
\mathscr{D} (\tau )=\sqrt{\left| \mathscr{A} +2\mathscr{B} \right|},
\]
with
\[
\begin{aligned}\mathscr{A}= & C_{i,i_{+},j,j_{+}}  +C_{j,j_{+},k,k_{+}}+C_{i,i_{+},l,l_{+}}+C_{k,k_{+},l,l_{+}}\\
= & \frac{A_{\perp}^{4}\tau^{4}}{15}\left[8+\cos2\varphi_{1}+\cos2\varphi_{2}+\cos2\left(\varphi_{1}+\varphi_{2}+\varphi_{3}\right)+\cos2\varphi_{3}\right]+O(\tau^{5}),\\
\mathscr{B}= & C_{i,j,j_{+},k}-C_{i,k,l,l_{+}}+C_{i,i_{+},j,l}-C_{j,k,k_{+},l}\\
= & \frac{2A_{\perp}^{4}\tau^{4}}{15}\left(\sin\varphi_{1}-\sin\varphi_{3}\right)\left[\sin\varphi_{2}-\sin(\varphi_{1}+\varphi_{2}+\varphi_{3})\right]+O(\tau^{5}).
\end{aligned}
\]
As a result, the WSL of $\mathscr{V}(\tau)$ is calculated to be 
Eq.~(13) in main text.

\subsection{Signal-to-noise ratio}

Here we estimate the data acquisition time $T$ required to observe the violation of the inequality above a certain signal-to-noise ratio for the quantum spin model. The model can be realized for example in the nitrogen-vacancy (NV) center in diamond with the measurement apparatus being the NV center electron spin and the target being a single $^{13}$C nuclear spin. The correlations are obtained using the photon counts. In this case, the direct observable is the
deviations of photon numbers as defined in Ref.~[\onlinecite{PfenderNC2019}] and [\onlinecite{JonasNC2022}]
\[
\tilde{C}_{ijkl}=\langle\delta n_{i}\delta n_{j}\delta n_{k}\delta n_{l}\rangle,
\]
where $\delta n_{i}=n_{i}-\langle n_{i}\rangle$ with 
$\langle n_{i}\rangle$ being the expectation value of the photon count at time $t_i$. The photon count
conditioned on the electron spin state $u$ is assumed to have the Poisson distribution
\[
p(n|\pm)=e^{-n_{\pm}}\frac{n_{\pm}^{n}}{n!},
\]
where $n_{\pm}$ be the average photon number collected in eachshot of measurement
when the electron spin is in the state $|u=\pm\rangle$. Then we have the photon number
distribution\cite{PfenderNC2019}
\[
p(n)=p(n|+)p(+)+p(n|-)p(-)
\]
and the averaged photon number\cite{JonasNC2022}
\[
\langle n\rangle=n_{+}p(+)+n_{-}p(-)=\frac{n_{+}+n_{-}}{2}+\frac{n_{+}-n_{-}}{2}\langle u\rangle.
\]
Considering the coefficient $(n_{+}-n_{-})/2$, the correlations of the
photon numbers 
\[
\begin{gathered}\tilde{C}_{ij} =\left(\frac{n_{+}-n_{-}}{2}\right)^{2}C_{ij},\\
\tilde{C}_{ijkl}  =\left(\frac{n_{+}-n_{-}}{2}\right)^{4}C_{ijkl}.
\end{gathered}
\]

Then we estimate the uncertainties of the second- and fourth-order correlations. Since the correlations are very weak and hence the photon
counts in different shots of measurements are nearly independent\cite{JonasNC2022,PfenderNC2019},
we have
\[
\begin{gathered}\delta\tilde{C}_{2}=\frac{\Delta^{2}}{\sqrt{M}},\\
\delta\tilde{C}_{4}=\frac{\Delta^{4}}{\sqrt{M}},
\end{gathered}
\]
where $M$ is the
total number of measurement shots and the fluctuation of the photon count in
each single shot of measurement is
\[
\Delta=\sqrt{\langle n^{2}\rangle-\langle n\rangle^{2}}=\sqrt{\frac{(n_{+}-n_{-})^{2}+2\left(n_{+}+n_{-}\right)}{4}}.
\]
Here we have used $\langle u\rangle=0$ in the quantum spin model.
Let $\gamma$ be the photon emission rate, $\eta$ be the photon
collection efficiency, and $\chi_{\rm ph}=\eta\gamma$ be the photon count rate. If the fluorescence contrast is set to be perfect\cite{RobledoNature2011},
$n_{+}=\chi_{\rm ph}\tau$ and $n_{-}=0$. Therefore,
\[
\Delta=\sqrt{\frac{(\chi_{\rm ph}\tau)^{2}+2\chi_{\rm ph}\tau}{4}}.
\]
Taking $T$ as the total data acquisition time and $M=T/\tau$ for sequential
measurement, we get
\[
\begin{gathered}\delta C_{2}=\left(\frac{2}{n_{+}-n_{-}}\right)^{2}\frac{\Delta^{2}}{\sqrt{M}}=\left(1+\frac{2}{\chi_{\rm ph}\tau}\right)\left(\frac{\tau}{T}\right)^{1/2},\\
\delta C_{4}=\left(\frac{2}{n_{+}-n_{-}}\right)^{4}\frac{\Delta^{4}}{\sqrt{M}}=\left(1+\frac{2}{\chi_{\rm ph}\tau}\right)^{2}\left(\frac{\tau}{T}\right)^{1/2}.
\end{gathered}
\]
For short $\tau$ (i.e., $\chi_{\rm ph}\tau\ll 1$), 
\[
\begin{gathered}\delta C_{2}\approx\left(\frac{2}{\chi_{\rm ph}\tau}\right)\left(\frac{\tau}{T}\right)^{1/2},\\
\delta C_{4}\approx\left(\frac{2}{\chi_{\rm ph}\tau}\right)^{2}\left(\frac{\tau}{T}\right)^{1/2}.
\end{gathered}
\]

Finally we estimate the signal-to-noise ratio for measuring $\mathscr{V}=\mathscr{N}/\mathscr{D}$ (here
we have neglected the dependence on $\tau$). Using the error propagation
formula, we get
\[
\delta\mathscr{V}=\sqrt{\left(\frac{1}{\mathscr{D}}\delta\mathscr{N}\right)^{2}+\left(\frac{\mathscr{N}}{\mathscr{D}^{2}}\delta\mathscr{D}\right)^{2}}
\]
where 
\begin{align}
\delta\mathscr{N} & \approx\sqrt{4}\delta C_{2}  =\frac{4}{\left(\sqrt{\chi_{\rm ph}\tau}\right)^{2}}\left(\frac{\tau}{T}\right)^{1/2}, \nonumber \\
\delta\mathscr{D} & \approx\frac{\delta(\mathscr{A}+2\mathscr{B})}{2\left(\mathscr{A}+2\mathscr{B}\right)^{1/2}}=\frac{1}{\mathscr{D}}\sqrt{3}\frac{4}{\left(\sqrt{\chi_{\rm ph}\tau}\right)^{4}}\left(\frac{\tau}{T}\right)^{1/2}. \nonumber
\end{align}
Using these errors, the error of $\mathscr{V}$ is estimated to be
\[
\begin{gathered}\delta\mathscr{V}=\left(\frac{\tau}{T}\right)^{1/2}\frac{1}{\left(\sqrt{\chi_{\rm ph}\tau}\right)^{2}}\sqrt{\left(\frac{4}{\mathscr{D}}\right)^{2}+\left(\frac{4\sqrt{3}\mathscr{N}}{\mathscr{D}^{3}}\frac{1}{\chi_{\rm ph}\tau}\right)^{2}}\\
=\left(\frac{\tau}{T}\right)^{1/2}\frac{1}{\left(\sqrt{\chi_{\rm ph}\tau}\right)^{2}}\sqrt{16\left(\frac{1}{\mathscr{D}}\right)^{2}+48\left(\mathscr{V}\frac{1}{\mathscr{D}^{2}}\frac{1}{\chi_{\rm ph}\tau}\right)^{2}}.
\end{gathered}
\]
Since $\mathscr{D}\approx2A_{\perp}^{2}\tau^{2}$ and $\mathscr{V}\approx1.25$,
the noise is dominated by the second term and hence we have
\[
\delta\mathscr{V}\approx2\left(\frac{\tau}{T}\right)^{1/2}\left(\frac{1}{\chi_{\rm ph}\tau}\right)^{2}\left(\frac{1}{A_{\perp}^{4}\tau^{4}}\right).
\]
To break the inequality by $R$ standard deviations, we need 
\[
\delta\mathscr{V}\le \frac{1}{4R},
\]
which requires the total data acquisition time
\[
T\gtrsim 64 R^{2}\tau\left(\frac{1}{\chi_{\rm ph}\tau}\right)^{4}\frac{1}{A_{\perp}^{8}\tau^{8}}=\frac{64}{\gamma} \frac{1}{\eta^{4}}\left(\frac{A_{\perp}}{\gamma}\right)^{3}\frac{R^2}{\left(A_{\perp}\tau\right)^{11}}.
\]
For parameters $A_{\perp}=2\pi\times 2.3~\mathrm{\mu sec}^{-1}$, $\omega=2\pi~\mathrm{\mu sec}^{-1}$,
$\gamma=1/(10~\mathrm{ns})$, $\eta=0.03$ and $\tau/T_{p}=0.023$,
we have $A_{\perp}\tau\approx 0.33$ and hence $T\approx 3$ hours for $R=5$.

\bibliographystyle{apsrev4-1}

%

